%% file: MLMC_on_magnetics.tex
\documentclass[journal,transmag]{IEEEtran}

\usepackage{amsthm}

\usepackage{amsmath}
\usepackage[cmintegrals]{newtxmath}
\usepackage{tikz,pgfplots}
\usepackage{enumitem}
\pgfplotsset{compat=newest}
\usepackage{booktabs}
\usepackage{siunitx}
\usepackage{multicol}
\usepackage[font=small]{caption}
\usepackage{subcaption}
\usepgfplotslibrary{fillbetween}
\usetikzlibrary{patterns}

\newcommand{\highlight}[1]{\textcolor{red}{#1}}

\newcommand{\mean}{\mathbb{E}} 
\newcommand{\meanMC}{\mathbb{E}_\mathrm{MC}} 
\newcommand{\meanML}{\mathbb{E}_\mathrm{ML}} 
\newcommand{\var}{\mathbb{V}} 
\newcommand{\richard}[1]{\hat{#1}} 
\newcommand{\randV}{\theta} 
\newcommand{\realW}{w} 
\newcommand{\srhs}{\text{s}}
\newcommand{\sstat}{\text{e}}
\newcommand{\dof}{\ell}
\newcommand{\ra}[1]{\renewcommand{\arraystretch}{#1}}
\renewcommand\vec{\mathbf}

\usepackage{eso-pic}
\AddToShipoutPicture*{\footnotesize\sffamily\raisebox{0.7cm}{\hspace{1.65cm}\fbox{\parbox{\textwidth}{\copyright~2019
IEEE. Personal use of this material is permitted. Permission from
IEEE must be obtained for all other uses, in any current or future
media, including reprinting/republishing this material for
advertising or promotional purposes, creating new collective works,
for resale or redistribution to servers or lists, or reuse of any
copyrighted component of this work in other works.}}}}

\begin{document}
\title{A multilevel Monte Carlo method for high-dimensional uncertainty quantification of low-frequency
electromagnetic devices}

\author{\IEEEauthorblockN{Armin Galetzka\IEEEauthorrefmark{1},
Zeger Bontinck\IEEEauthorrefmark{1,2},
Ulrich R\"omer\IEEEauthorrefmark{3}, and
Sebastian Sch\"ops\IEEEauthorrefmark{1,2}}
\IEEEauthorblockA{\IEEEauthorrefmark{1}Institut f\"ur Theorie Elektromagnetischer Felder, Technische Universit\"at Darmstadt, Darmstadt 64289, Germany}
\IEEEauthorblockA{\IEEEauthorrefmark{2}Graduate School of Computational Engineering, Technische Universit\"at Darmstadt, Darmstadt 64293, Germany}
\IEEEauthorblockA{\IEEEauthorrefmark{3}Institut f\"ur Dynamik und Schwingungen, Technische Universit\"at Braunschweig 38106, Germany}
\thanks{Manuscript received August 3, 2018; revised November 9, 2018 and accepted April 1, 2019. Date of publication May 3, 2019; date of current version July 18, 2019. Corresponding author: S. Schöps (e-mail: schoeps@temf.tu-darmstadt.de). Color versions of one or more of the figures in this paper are available online at http://ieeexplore.ieee.org. Digital Object Identifier 10.1109/TMAG.2019.2911053.}}

\markboth{IEEE TRANSACTIONS ON MAGNETICS, VOL. 55, NO. 8, August 2019}{}
%



\IEEEtitleabstractindextext{%
\begin{abstract}
This work addresses uncertainty quantification of electromagnetic devices determined by the eddy current problem. The multilevel Monte Carlo (MLMC) method is used for the treatment of uncertain parameters while the devices are discretized in space by the finite element method. Both methods yield numerical approximations such that the total errors is split into stochastic and spatial contributions. We propose a particular implementation where the spatial error is controlled based on a Richardson-extrapolation-based error indicator. The stochastic error in turn is efficiently reduced in the MLMC approach by distributing the samples on multiple grids. The method is applied to a toy problem with closed-form solution and a permanent magnet synchronous machine with uncertainties. The uncertainties under consideration are related to the material properties in the stator and the magnets in the rotor. The examples show that the error indicator works reliably, the meshes used for the different levels do not have to be nested and, most importantly, MLMC reduces the computational cost by at least one order of magnitude compared to standard Monte Carlo.
\end{abstract}

}

\maketitle

\IEEEdisplaynontitleabstractindextext

%
\IEEEpeerreviewmaketitle

\input{Introduction}
\input{Methodology}
\input{Application}
\input{Results}
\input{Conclusion}

\section*{Acknowledgment}
This work is supported by the German BMBF in the context of the SIMUROM project (05M2013) and the PASIROM project (05M18RDA), by the DFG (SCHO1562/3-1), by the 'Excellence Initiative' of the German Federal and State Governments and the Graduate School of CE at TU Darmstadt (GSC 233/2).

\ifCLASSOPTIONcaptionsoff
  \newpage
\fi

\end{document}

%% file: Introduction.tex

\section{Introduction}
%
%
%
%
\IEEEPARstart{T}{he} 
consideration of uncertainties in modeling and simulation becomes increasingly popular in electrical engineering applications. In real-world problems, the number of uncertain random parameters is often very large, e.g. due to variations in the material, geometry and sources \cite{Clenet_2013aa}. In a magnetics context, uncertainties in the material have been addressed for example in~\cite{Ramarotafika_2012ab,Roemer_2016aa,Jankoski2017}. An uncertain material geometry was considered in~ \cite{Loukrezis_2017aa}, whereas uncertainties in sources have been discussed in~\cite{Offermann_2013aa}. Often one is interested in the forward propagation of those uncertainties to quantify the yield, rates of failure, stochastic moments, e.g. the mean value and sensitivities. To this end, one is concerned with the solution of partial differential equations with random inputs, e.g. stemming from Maxwell's equations.

There are several approaches to deal with uncertainties. Stochastic collocation, stochastic Galerkin, often in combination with polynomial chaos methods, see e.g. \cite{ref:xiu}, belong to a class of methods that are very efficient if the number of uncertain input parameters is small. As the dimensionality of the uncertainty increases, those methods become inefficient and eventually unaffordable, due to the \emph{curse-of-dimensionality}. Sparse grids \cite{Bungartz_2004} can be used in this case, but the curse-of-dimensionality will only be mitigated. Additionally, if the solution is not smooth with respect to the random inputs, those methods are not well suited.

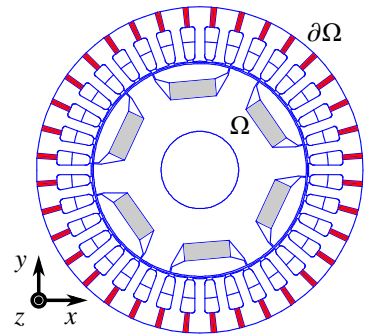
\begin{figure}[ht]
		\centering
		\def\svgwidth{0.5\columnwidth}\input{pmsm_coarse.pdf_tex}
		\caption{Six pole permanent magnet synchronous machine. The gray regions depict permanent magnets, red regions correspond to welding regions.}
		\label{fig:pmsm}
		\vspace{-0.8em}
\end{figure}

A classical way to deal with such problems is the Monte Carlo (MC) method, since its convergence rate is independent of the number of uncertain inputs. However, the convergence rate is $\mathcal{O}(N^{-1/2})$, where $N$ is the number of drawn samples, which is slow compared to the methods mentioned above. Therefore, the computational cost becomes prohibitive for high accuracies, since the underlying partial differential equation (PDE) needs to be solved for every sample.

A variety of so-called variance reduction techniques can be used to improve the convergence rate of MC. One of them is the well known quasi Monte Carlo (QMC) approach. Instead of choosing the samples randomly and independently, the samples are chosen out of a sequence, resulting in an error convergence rate of $\mathcal{O}(\log(N)^{M}/N)$, which suffers from the curse of dimensionality as well if the number of random variables $M$ is large \cite{ref:QMC}. For a moderate number of samples $N$ and a moderate $M$, the convergence is given approximately by $\mathcal{O}(N^{-1})$~\cite{Gunzburger_2015aa}. 

Another popular technique for variance reduction is the multilevel Monte Carlo (MLMC) method. The technique was first introduced by Heinrich \cite{ref:Heinrich} and then extended by Giles \cite{ref:Giles_Path_Sim, ref:Giles_2015aa}.
Recent works deal with particular partial differential equations with random inputs \cite{ref:Barth_MLMC_PDE, ref:cliffe} or stochastic parameters with a lack of regularity \cite{ref:Teckentrup_FEM}. Another focus is the combination of MLMC with QMC to further reduce the complexity \cite{ref:QMLMC_waterhouse, ref:QMLMC_kuo}.

The aim of this work is to investigate the applicability of the MLMC method to real world problems from electrical engineering, while prior works, e.g. \cite{Galetzka_2017aa}, were concerned with toy problems only. The scheme is adopted for the simulation of a permanent magnet synchronous machine (PMSM), see Fig.~\ref{fig:pmsm}. Different causes of uncertainty have already been studied in literature, e.g. stator-teeth length~\cite{Offermann_2015aa}, rotor eccentricity~\cite{Bontinck_2016aa} and non-linear material behaviour~\cite{Roemer_2014aa}. In this work we consider uncertainties in the permanent magnets, see e.g.~\cite{Jurisch_2007aa}. Since there is a trend to use segmentation for the construction of the stator, we also consider uncertainties in the welding regions between the different stator teeth. It is known that welding affects the permeability of the steel used for constructing the machine~\cite{Clerc_2012aa}. Due to the fact that the uncertainties destroy the symmetry of the machine, one has to model the full machine. Furthermore, the number of random parameters is large, thus the MLMC method is a proper approach for this problem. The quantity of interest is the mean magnetic energy, without any current excitation. To calculate this quantity up to a desired accuracy, an appropriate error indicator has to be used.

The outline of this paper is as follows: in Section \ref{sec:methodology}, a general magnetoquasistatic PDE is formulated together with the finite element (FE) scheme. In addition, the MLMC approach and the Richardson extrapolation-based error indicator are introduced. In Section \ref{sec:application}, the applications are introduced. Firstly, the approach is applied to a coaxial cable with uncertainties. This enables us to verify the results obtained by the extrapolation. The second application is a PMSM. The numerical results can be found in Section \ref{sec:results}. Finally, in Section~\ref{sec:conclusion}, conclusions are drawn.

%% file: pmsm_coarse.pdf_tex
\begingroup%
  \makeatletter%
  \providecommand\color[2][]{%
    \errmessage{(Inkscape) Color is used for the text in Inkscape, but the package 'color.sty' is not loaded}%
    \renewcommand\color[2][]{}%
  }%
  \providecommand\transparent[1]{%
    \errmessage{(Inkscape) Transparency is used (non-zero) for the text in Inkscape, but the package 'transparent.sty' is not loaded}%
    \renewcommand\transparent[1]{}%
  }%
  \providecommand\rotatebox[2]{#2}%
  \ifx\svgwidth\undefined%
    \setlength{\unitlength}{349.96875bp}%
    \ifx\svgscale\undefined%
      \relax%
    \else%
      \setlength{\unitlength}{\unitlength * \real{\svgscale}}%
    \fi%
  \else%
    \setlength{\unitlength}{\svgwidth}%
  \fi%
  \global\let\svgwidth\undefined%
  \global\let\svgscale\undefined%
  \makeatother%
  \begin{picture}(1,1)%
    \put(0,0){\includegraphics[width=\unitlength]{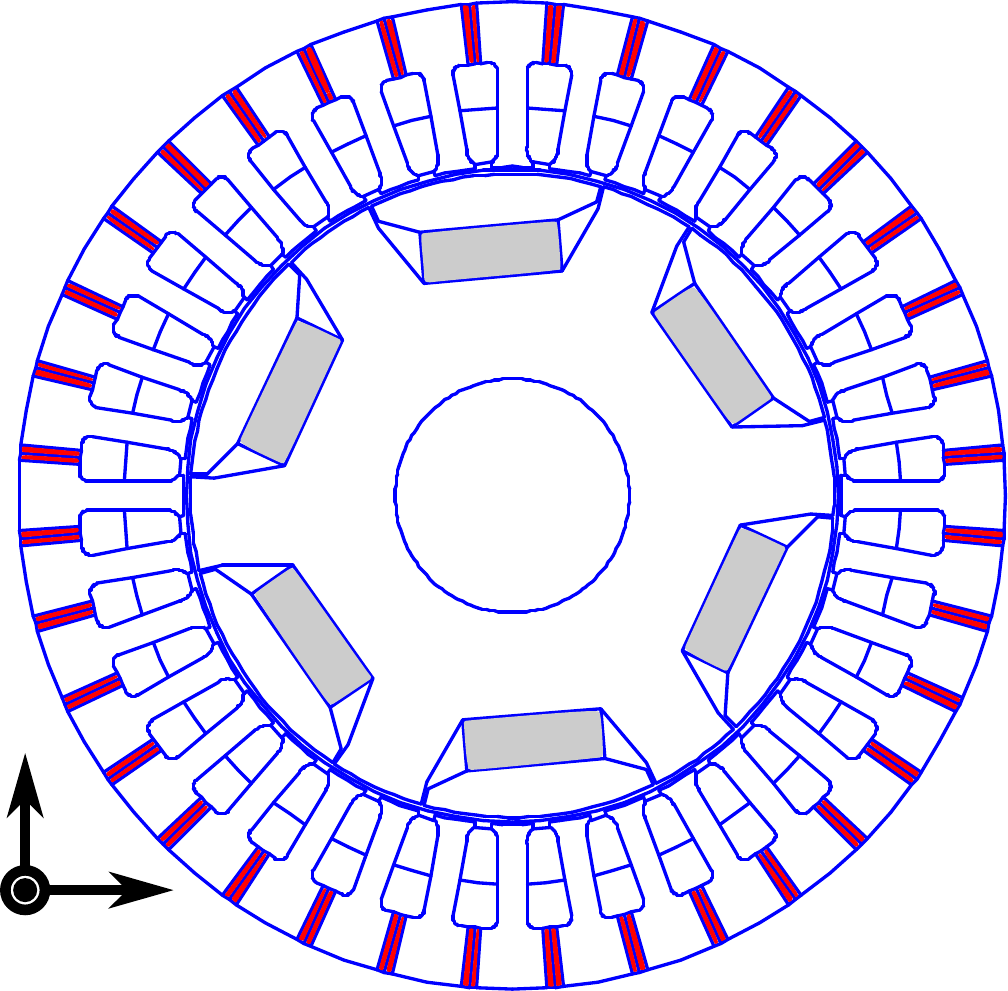}}%
    \put(0.83,0.88){\color[rgb]{0,0,0}\makebox(0,0)[lb]{\smash{$\partial\Omega$}}}%
    \put(0.6,0.6){\color[rgb]{0,0,0}\makebox(0,0)[lb]{\smash{$\Omega$}}}%
    \put(0.1,0.03){\color[rgb]{0,0,0}\makebox(0,0)[lb]{\smash{$x$}}}%
    \put(-0.05,0.2){\color[rgb]{0,0,0}\makebox(0,0)[lb]{\smash{$y$}}}%
    \put(-0.05,0.03){\color[rgb]{0,0,0}\makebox(0,0)[lb]{\smash{$z$}}}%
  \end{picture}%
\endgroup%

%% file: Methodology.tex
\section{Methodology and Theory}
\label{sec:methodology}
The underlying equations are introduced first, followed by an introduction to the MLMC method and theory. After that, the Richardson extrapolation is introduced and discussed.
\subsection{Magnetoquasistatic formulation and discretization}
We consider devices that operate at low frequencies and treat them as magnetoquasistatic, i.e. we disregard the displacement current density $\|\imath\omega\vec{D}\|<<\|\vec{J}_\srhs\|$ with respect to the total current density, where $\omega$ denotes the angular frequency of a sinusoidal excitation. Using the $A^\star$ formulation \cite{Emson_1988aa} in the frequency domain one obtains the following PDE
\begin{align}
	\imath\sigma(\vec{x})\omega \vec{A}(\vec{x})
	+
	\nabla \times \left( \nu(\vec{x}) \nabla \times \vec{A}(\vec{x}) \right)
	&= 	\vec{J}_\srhs(\vec{x}) &&\vec{x}\  \text{in }\Omega,
	\label{eq:magquasistat}\\
	\vec{A}_\text{t}(\vec{x}) &= 0  &&\vec{x}\ \text{on }\partial\Omega,
\intertext{with imposed Dirichlet boundary conditions, where $\nu=\mu^{-1}$ is the reluctivity or inverse permeability, $\sigma$ is the electrical conductivity, $\vec{J}_\srhs$ is the source current density, $\vec{A}$ is the magnetic vector potential, $\vec{A}_\text{t}$ denotes the tangential vector components of $\vec{A}$ and $\Omega$ refers to the computational domain.
In the static regime $\omega=0$, we obtain the special case}
	\nabla \times \left( \nu(\vec{x}) \nabla \times \vec{A}(\vec{x}) \right)  &= \vec{J}_\srhs(\vec{x})&&\vec{x}\  \text{in }\Omega,
	\label{eq:magstatic}
\end{align}
with the same boundary condition. In either case, the total source current $\vec{J}_\srhs$ is given by
\begin{equation}
\vec{J}_\srhs(\vec{x})= \vec{J}_\sstat(\vec{x}) - \nabla \times \left( \nu_\text{pm}(\vec{x})\vec{B}_\text{r}(\vec{x})\right),
\end{equation}
where $\vec{J}_\sstat$ is an impressed external current density, $\vec{B}_\text{r}$ is the remanent magnetic field and $\nu_\text{pm}$ is the reluctivity of the permanent magnets. 

The quantity of interest (QoI) can be any functional $F(\vec{A})$, for example the magnetic energy
\begin{equation}
E = \frac{1}{2}\int_{\Omega}\nu\vec{B}\cdot\vec{B}^*\;\text{d}\Omega,
\label{eq:magneticEnergyGeneral}
\end{equation}
with $\vec{B}=\nabla\times\vec{A}$ and $*$ denoting the complex conjugate.

\begin{figure}
	\vspace{-0.1em}
	\begin{subfigure}[c]{0.5\linewidth}
		\centering
		\includegraphics[width=\linewidth]{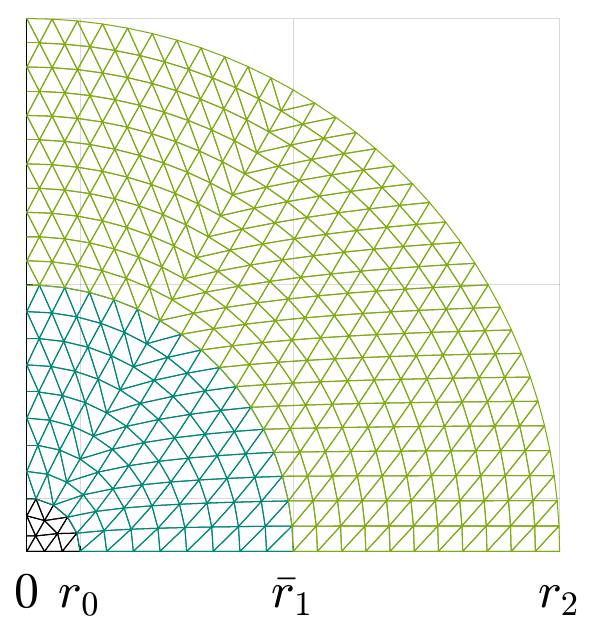}
		\caption{Coarsest level $\ell=0$.}
		\label{fig:coax_mesh_coarse}
	\end{subfigure}%
	\begin{subfigure}[c]{0.5\linewidth}    
		\centering
		\includegraphics[width=\linewidth]{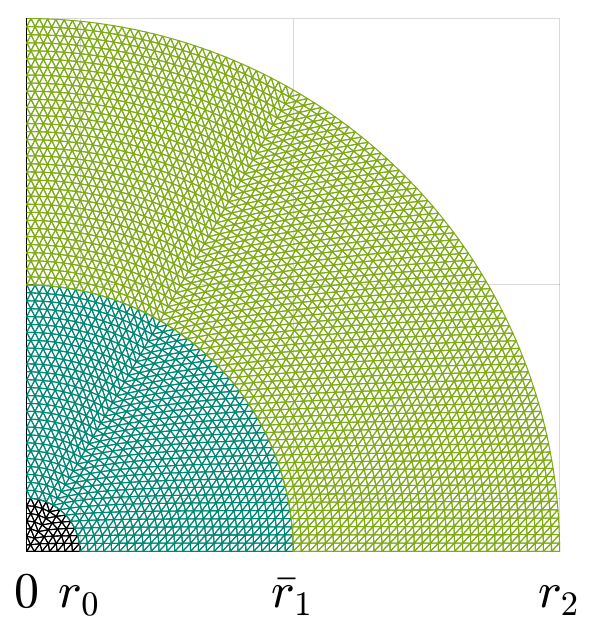}
		\caption{Finest level $\ell=L$.}
		\label{fig:coax_mesh_fine}
	\end{subfigure}
	\caption{Mesh for a toy example (coaxial cable).}
	\vspace{-0.8em}
\end{figure}

\label{key}
In order to solve the above mentioned PDEs, the finite element (FE) method is used \cite{ref:Monk}. In particular, the Galerkin approach is applied, where ansatz and test functions are chosen identically. The magnetic vector potential is approximated as
\begin{equation}
\vec{A}_\ell(\vec{x})=\sum_{i=1}^{n_{\dof}} a_i \vec{w}_i(\vec{x}),
\end{equation}
whereat $a_i$ are the degrees of freedom (DoF), $\vec{w}_i$ are vectorial basis functions defined on a triangularization of $\Omega$ and $n_{\dof}$ represents the level-dependent number of DoFs. We denote by ${\ell\in[0,\dots,L]}$ the level of the mesh. A large $\ell$ corresponds to a fine mesh and to an accurate solution, which however, also requires high computational costs.
We define the mesh size $h_\ell$ as the maximum edge length in the mesh of level $\ell$. Examples of meshes of different resolution are depicted in Figures \ref{fig:coax_mesh_coarse} and \ref{fig:coax_mesh_fine}.

In this work, lowest order ansatz and test functions are employed. When considering planar 2D problems, the edge shape functions only have a $z$-component and can be constructed from the nodal shape functions $N_i(\vec{x})$ as follows
\begin{equation}
	\label{eq:discrete_curlcurl}
	\vec{w}_i(\vec{x})=\frac{N_i(\vec{x})}{l_z}\vec{e}_z,
\end{equation}
where $l_z$ and $\vec{e}_z$ refer to the length of the device and the unit vector in the $z$-direction, respectively. This leads to the system of equations 
\begin{equation}
\mathbf{K}_\nu\mathbf{a}+\imath\omega\mathbf{M}_\sigma\mathbf{a}=\mathbf{j}_\srhs,
\label{eq:magnetoquasi_discret}
\end{equation}
where
\begin{subequations}
\begin{equation}
K_{\nu,i,j}=\int_{\Omega}\nu(\vec{x})\nabla\times \vec{w}_j(\vec{x})\cdot\nabla\times \vec{w}_i(\vec{x})\;\mathrm{d}V,
\end{equation}
\begin{equation}
M_{\sigma,i,j}=\int_{\Omega} \sigma(\vec{x})\vec{w}_j(\vec{x})\cdot\vec{w}_i(\vec{x})\;\mathrm{d}V,
\end{equation}
and
\begin{equation}
j_{\srhs,i}= \int_{\Omega}  \left(\vec{J}_\sstat(\vec{x})\cdot\vec{w}_i(\vec{x})- \vec{H}_\text{pm}(\vec{x})\cdot \nabla \times \vec{w}_i(\vec{x}) \right)\;\mathrm{d}V
\end{equation}
\end{subequations}
which is essentially a Poisson equation with a non-standard right-hand-side because the permanent magnets introduce a singular excitation which may affect the convergence order of the FE method. 
However, this type of problem is well understood e.g. \cite{Babuska_2005ab,Cohen_2011aa}.

\subsection{Multilevel Monte Carlo Method}
In the following we introduce the MLMC method. We are interested in the solution of the elliptic PDEs \eqref{eq:magquasistat} and~\eqref{eq:magstatic} with random input parameters. 
{Let us introduce} the probability space $\left(\Theta, \Sigma, P\right)$, where $\Theta$ is the space of possible outcomes, $\Sigma$ the set of events and $P$ the probability measure. 
The random PDE reads
\begin{equation}
{\nabla \times \left( \nu(\vec{x},\theta) \nabla \times \vec{A}(\vec{x},\theta) \right) = \vec{J}_\srhs(\vec{x},\theta), \quad \vec{x}\  \text{in }\Omega,}
\label{eq:general_SPDE}
\end{equation} 
which is assumed to hold with probability one, where the solution {$\vec{A}(\vec{x},\theta)$} inherits the stochastic nature. It should be noted, that a random reluctivity $\nu$ represents both uncertainties in the material parameters and the geometry at material interfaces. The QoI can be the solution itself or a functional {$F(A)$}. In the following, the QoI is denoted as {$W(\theta)=F\left(\vec{A}(\cdot,\theta)\right)$}, which is again a random variable $W: \Theta \to \mathbb{R}$. The purpose is to estimate statistical measures of the QoI. Our focus will be on stochastic moments, in particular, on the mean value $\mean[W]$. It should be emphasized that an approximation of the mean value operator can be used to numerically compute higher order moments and failure probabilities as well. Moreover, the probability density function of $W$ can be approximated based on a MLMC approximation of the mean value using the maximum entropy method \cite{Bierig_2016}.

The following exposition can be found in \cite{ref:Giles_2015aa}. The main identity of the MLMC method reads 
\begin{equation}
\mean[W_L]= \mean[W_0]+\sum_{\ell=1}^L \mean[W_\ell-W_{\ell-1}],
\label{eq:MLMC_main}
\end{equation}
which can be derived by using the linearity of $\mean[\cdot]$ and by adding $\mean[W_L]$ on both sides of
$${\mean[W_0]-\mean[W_0]+\cdots+\mean[W_{L-1}]-\mean[W_{L-1}]=0}.$$ 
The solutions $W_0 \dots W_L$ refer to FE approximation on mesh level $\ell$.
The mean values in \eqref{eq:MLMC_main} are evaluated with the conventional MC method, which leads to
\begin{align}
\mean[W_L] &\approx \meanMC[W_0]+\sum_{\ell=1}^L \meanMC[W_\ell-W_{\ell-1}] \label{eq:MCapprox}\\
&= N_0^{-1} \sum_{i=1}^{N_{0}} W_0^{(i,0)} + \sum_{\ell=1}^L N_\ell^{-1} \sum_{i=1}^{N_\ell} \left( W_\ell^{(i,\ell)} - W_{\ell-1}^{(i,\ell)} \right) \notag \\
&= \meanML[W_L] \notag
\end{align} 
where $\meanML[\cdot]$ represents the multi level estimator, whereas $\meanMC[\cdot]$ represents the unbiased MC estimator 
\begin{equation}
\meanMC[W]=N^{-1} \sum_{i=1}^N W^{(i)}.
\end{equation}
Note that the superscript $(i,\ell)$ expresses the fact that in the inner summation $W_\ell^{(i,\ell)}$ and $W_{\ell-1}^{(i,\ell)}$ are evaluated using the same sample with index $i$. Also, a new sampling set is considered for each term of the outer summation, reflected by the superscript $\ell$. The MC estimator is an unbiased estimator with underlying properties
\begin{align}
\mean[\meanMC[W]] &= \mean[W], \\
\var[\meanMC[W]] &= N^{-1} \var[W],
\end{align}
where the variance is defined as $\var[X]=\mean\left[\left(X-\mean[X]\right)^2\right]$. This implies, that the variance of our MLMC approximation \eqref{eq:MCapprox} is given by
\begin{align}
\var\left[\meanML[W_L]\right] &= \var\left[\meanMC[W_0]+\sum_{\ell=1}^L \meanMC[W_\ell-W_{\ell-1}]\right]  \\
&= N_0^{-1} \underbrace{\var[W_0]}_{=V_0} + \sum_{\ell=1}^L N_{\ell}^{-1} \underbrace{\var[W_\ell-W_{\ell-1}]}_{=V_\ell}.
\label{eq:totalVar}
\end{align}
Since $W^{(i,\ell)}_{\ell}$ and $W^{(i,\ell)}_{\ell-1}$ approximate the same problem and use the same samples, their difference is small and so is their variance. We can already assume that the variance on the coarsest, computational cheapest, level is dominating. 
Thereby, the question which arises is how to choose $N_0$, respectively $N_\ell$. The best approximation would be obtained by sampling only the highest level, but this would also imply a very high computational cost. Instead, in the MLMC method, the samples are distributed onto the different levels using a cost-benefit consideration. To have a more quantitative measure, the total cost is introduced by
\begin{equation}
C=\sum_{\ell=0}^L N_\ell C_\ell,
\label{eq:costsGeneral}
\end{equation}
where $C_\ell$ denotes the cost for one sample on level $\ell$. After minimizing \eqref{eq:totalVar}, under the constraint of a fixed computational budget \eqref{eq:costsGeneral}, with a Lagrange multiplier method \cite{ref:Giles_2015aa}, we obtain the optimal number of samples per level as
\begin{equation}
N_\ell = \varepsilon^{-2} \sqrt{V_\ell/C_\ell}\sum_{\ell=0}^L \sqrt{V_\ell C_\ell},
\label{eq:NlBasic}
\end{equation}
where $\varepsilon$ is a user specified accuracy. 

A common measure of the error, resulting from the MLMC and the FE approximation is the mean square error (MSE), defined as
\begin{align}
\textrm{MSE} &= \mean\left[\left( \meanML[W_L]-\mean[W]\right)^2\right] \notag \\
&= \var[\meanML[W_L]] + \left(\mean[\meanML[W_L]-W]\right)^2 \notag \\
&= \var[\meanML[W_L]] + \left(\mean[W_L-W]\right)^2.
\label{eq:MSE}
\end{align}
Hence, the MSE is divided into a stochastic error which is equal to the variance of the MLMC estimator and a so-called weak error given by $\left(\mean[W_L-W]\right)^2$, which is related to the FE error. The stochastic and weak error can be reduced by increasing the number of samples and by using meshes with a finer resolution, respectively.
%
%
If the variance and the weak error $\left(\mean[W_L-W]\right)^2$ are smaller than $\varepsilon^2/2$, the MSE can be kept below a bound of $\varepsilon^2$. This can be achieved, by fulfilling the conditions of the following theorem:
\newtheorem{mlmcTheorem}{Theorem}
\begin{mlmcTheorem}[\cite{ref:Teckentrup_FEM}]
\label{th:mlmcTheorem}
Let $W$ denote a random variable, and let $W_\ell$ denote the corresponding level $\ell$ numerical approximation.
If there exist independent unbiased estimators $\meanMC[W_0]$, $\meanMC[W_\ell-W_{\ell-1}]$ based on $N_\ell$ Monte Carlo samples, each with expected cost $C_\ell$ and variance $V_\ell$, and positive constants $\alpha,\beta,\gamma,c_1,c_2,c_3$ such that $\alpha \ge \frac{1}{2}\min(\beta,\gamma)$ and
\begin{enumerate}[label=(\roman*)]
\item $|\mean\left[W_\ell-W\right]| \le c_1 h_\ell^{\alpha}$ \label{eq:theorem1},
\item $V_\ell \le c_2 h_\ell^{\beta} $ \label{eq:theorem3},
\item $C_\ell \le c_3 h_\ell^{-\gamma}$ \label{eq:theorem4},
\end{enumerate}
then there exists a positive constant $c_4$ such that for any $0<\varepsilon < \mathrm{e}^{-1}$ there are values $L$ and $N_\ell$ for which the multilevel estimator
\begin{equation}
\meanML[W_L] = \meanMC[W_0] + \sum_{\ell=1}^L \meanMC[W_\ell-W_{\ell-1}]
\end{equation}
has a mean square error with bound
\begin{equation}
\textup{\textrm{MSE}}= \mean\left[(\meanML[W_L]-\mean[W])^2\right] < \varepsilon^2
\label{eq:MSETheorem}
\end{equation}
with a computational complexity $C$ with bound
\begin{equation}
C \le \left\{\begin{array}{ll} c_4\varepsilon^{-2} & \beta > \gamma, \\ c_4\varepsilon^{-2}(\log \varepsilon)^2 & \beta = \gamma, \\
         c_4\varepsilon^{-2-(\gamma-\beta)/\alpha} & \beta < \gamma. \end{array}\right.
\end{equation}
\end{mlmcTheorem}
The parameter $\alpha$ measures the weak error decay, whereas $\beta$ describes the decay of the variance. Both can be derived a priori for a given problem class, e.g. an elliptic model problem, see \cite{ref:Teckentrup_FEM}. If the analysis of the underlying problem is too complex, these constants can still be determined numerically in a pre-processing step, as discussed in \cite{ref:Giles_2015aa}. The constant $\gamma$ dictates the growth of the costs. If the mesh discretization is sufficiently fine, the cost for one sample is dominated by solving a linear system of equations with $n_{\dof}$ degrees of freedom. Since we solve 2D problems with lowest order basis functions, it holds that $n_{\dof} \approx h_\ell^{-2}$ where we for now disregard any special treatment of singularities on the right-hand side. We rewrite condition \ref{eq:theorem4} as
\begin{equation}
C_\ell \le \hat{c}_3 \left(\frac{1}{\sqrt{n_{\dof}}}\right)^{-\gamma}=\hat{c}_3 n_{\dof}^{\gamma/2}.
\end{equation}
In the case of a 2D problem with an optimized code and solver, it is possible to obtain $\gamma \approx 2$, see \cite{ref:cliffe}. To analyze the best possible efficiency gains by the MLMC method it will be assumed that $\gamma =2$ in the following, although the costs of the actual  implementation may be larger, e.g. due to a suboptimal linear solver. Hence, we define the total cost as
\begin{equation}
C=\sum_{\ell=0}^L N_\ell n_{\dof}.
\label{eq:costsGamma2}
\end{equation}

\subsection{Richardson extrapolation}
\label{sec:Rich_ex}
To keep the weak error below $\varepsilon^2/2$, the last level $L$ has to be sufficiently fine. Following \cite{ref:Giles_2015aa}, the choice of $L$ will be based on Richardson extrapolation in the following. Let $\realW=W(\theta)$ denote a random realization. Since the exact solution $\realW$ is unknown, we have to use an approximation. The FE error on level $\ell$ can be written as 
\begin{equation}
\realW=\realW_\ell+Kh_\ell^{k_0}+\mathcal{O}\left(h_\ell^{k_1}\right),
\label{eq:Taylor1}
\end{equation}
where $K$ is an unknown constant, $\realW_\ell$ is a realization of $W_\ell$ on level $\ell$, $k_0$ is the (known) convergence rate with $k_0 < k_1$. In the following we assume that
\begin{equation}
h_\ell=h_0\Delta^\ell
\label{eq:hLevel}
\end{equation}
holds, where $h_0$ is the initial mesh size on level $0$ and $\Delta \in (0,1)$ is the geometric refinement step. 
The FE error representation at a finer level $h_{\ell+1}$ leads to
\begin{align}
\realW&=\realW_{\ell+1}+Kh_{\ell+1}^{k_0}+\mathcal{O}\left(h_{\ell+1}^{k_1}\right), \notag \\
 &=\realW_{\ell+1}+Kh_\ell^{k_0}\Delta^{k_0}+\mathcal{O}\left(h_{\ell}^{k_1}\Delta^{k_1}\right), \notag \\
 &=\realW_{\ell+1}+Kh_\ell^{k_0}\Delta^{k_0}+\mathcal{O}\left(h_{\ell}^{k_1}\right).
 \label{eq:Taylor2}
\end{align}
Multiplying \eqref{eq:Taylor2} with $\Delta^{-k_0}$ and subtracting \eqref{eq:Taylor1} from \eqref{eq:Taylor2} yields the new approximation
\begin{equation}
\realW=\underbrace{\frac{\Delta^{-k_0}\realW_{\ell+1}-\realW_\ell}{\Delta^{-k_0}-1}}_{=\richard{\realW}_\ell}+\mathcal{O}\left(h_{\ell}^{k_1}\right).
\label{eq:richard}
\end{equation}
Hence, we have improved the convergence order from $k_0$ to $k_1$, compared to the conventional FE convergence. For sufficiently regular geometries, data and QoI there holds $k_0 = 2$ and $k_1 = 3$, i.e. a gain of one convergence order can be expected. The Richardson extrapolator $\richard{W}_\ell$ defined in~\eqref{eq:richard} can replace the exact solution $W$ to obtain an indicator of the weak error. 

Yet, computing $\mean\left[W_\ell - \richard{W}_\ell\right]^2$ is even more expensive than computing $\mean[W_\ell]$ and hence, another approximation is required. To this end we employ the so-called first-order second-moment method. {We assume in the following that the underlying problem depends on $M$ independent random input variables $\mathbf{Y}=(Y_1,Y_2,\dots,Y_M)$, such that $\mathbf{Y}: \Theta \to \Gamma \subset \mathbb{R}^M$. A random realization is again denoted by $\mathbf{y}=\mathbf{Y}(\theta)$. Let $f_\mathbf{Y}$ and $\boldsymbol{\mu_Y}$ denote the joint probability density function and the mean value of $\mathbf{Y}$, respectively}. By abuse of notation we write $W(\mathbf{y})$. The Taylor series of $W(\mathbf{y})$ around $\boldsymbol{\mu_Y}$ reads
\begin{align}
W(\mathbf{y}) &=W(\boldsymbol{\mu_Y})+ \sum_{i=1}^{M} \frac{\partial W(\boldsymbol{\mu_Y})}{\partial y_i}(y_i-\mu_{Y,i})\label{eq:FOSM_Taylor}  \\
&+ \frac{1}{2}\sum_{i=1}^{M}\sum_{j=1}^{M} \frac{\partial^2 W(\boldsymbol{\mu_Y})}{\partial y_i \partial y_j}(y_i-\mu_{Y,i})(y_j-\mu_{Y,j})+ \ldots \, . \notag
\end{align}
The first stochastic moment is defined as
\begin{equation}
\mean[W(\mathbf{Y})] = \int_{\Gamma} W(\mathbf{y}) f_\mathbf{Y}(\mathbf{y})\, \text{d}\mathbf{y},
\label{eq:FOSM_mean}
\end{equation}
which can be expressed, using all terms in the Taylor series  \eqref{eq:FOSM_Taylor} up to order one, as
\begin{align}
\mean[W(\mathbf{Y})] &\approx \underbrace{\int_{\Gamma} W(\boldsymbol{\mu_Y})f_\mathbf{Y}(\mathbf{y})\,\text{d}\mathbf{y}}_{=W(\boldsymbol{\mu_Y})} \nonumber \\ 
&+ \int_{\Gamma} \sum_{i=1}^{M} \frac{\partial W(\boldsymbol{\mu_Y})}{\partial y_i}(y_i-\mu_{Y,i})f_\mathbf{Y}(\mathbf{y})\,\text{d}\mathbf{y} \nonumber \\
&= W(\boldsymbol{\mu_Y}) + \sum_{i=1}^{M} \frac{\partial W(\boldsymbol{\mu_Y})}{\partial y_i} \underbrace{\int_{\Gamma} (y_i-\mu_{Y,i})f_\mathbf{Y}(\mathbf{y})\,\text{d}\mathbf{y}}_{=0} \nonumber \\
&= W(\boldsymbol{\mu_Y}). \label{eq:FOSM_approx}
\end{align}
Equation \eqref{eq:FOSM_approx} allows us to approximate the weak error at level $\ell$ as 
\[
\mean\left[W_\ell - \richard{W}_\ell \right]^2 \approx  \left|W_\ell(\boldsymbol{\mu_Y})) - \richard{W}_\ell(\boldsymbol{\mu_Y})) \right|^2
\]
which significantly reduces the computational effort. In particular, we can choose the number of levels a priori such that 
\begin{equation}
\left | W_L(\boldsymbol{\mu_Y})) - \richard{W}_L(\boldsymbol{\mu_Y})) \right|^2 \le \frac{\varepsilon^2}{2}.
\label{eq:RichardsonEstimator}
\end{equation}


%% file: Application.tex
\section{Application}
\label{sec:application}
The MLMC method is applied to two examples. First we consider an academic toy example, for which a closed form solution exists, which is used to verify the MLMC simulation with the Richardson error indicator. The influence of the mesh type is also analyzed. Finally, the method is applied to a PMSM.
\subsection{Coaxial cable}
\begin{figure}[t]
		\vspace{-0.5em}
		\centering
		\def\svgwidth{0.5\columnwidth}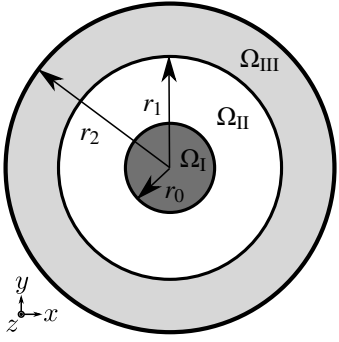
		\caption{Cross sectional view of a coaxial cable, where $\Omega_\text{I},\Omega_\text{II}$ and $\Omega_\text{III}$ represent the inner wire, the region with air and the outer steel pipe, respectively.}
		\label{fig:tube}
\end{figure}

Figure \ref{fig:tube} shows the cross-sectional view of a conducting wire surrounded by a pipe.
The geometry is split into three regions, such that the computational domain is given as $\overline{\Omega} = \overline{\Omega}_\text{I} \cup \overline{\Omega}_\text{II} \cup \overline{\Omega}_\text{III} $. Region $\Omega_\text{I}$ (dark grey) depicts the inner wire carrying the current $I_\text{s}$ and region $\Omega_\text{II}$ (white) is an area filled with air. These regions are modelled with vanishing conductivity and have the permeability of vacuum. The last region, i.e. $\Omega_\text{III}$ (light grey), depicts the outer pipe that is concentric to the inner wire. This outer pipe has a conductivity $\sigma_{\text{III}}$ and a permeability $\mu$. It is assumed that the radius $r_2$ is negligible compared to the length $L_\text{tube}$ of the coaxial cable, i.e $r_2 \ll L_\text{tube}$, thus boundary effects can be neglected. The current $I_\text{s}$ varies sinusoidally at an angular frequency $\omega$ and is given by
\begin{equation}
\vec{I}_{\text{s}}(r)=I_\text{s}(r)\sin{(\omega t)}\,\vec{e}_z, \quad r \leq r_0.
\end{equation}
In addition, it is assumed that the current is homogeneously distributed over the face, which leads to the source current density $\vec{J}_{\text{s}}=\vec{I}_\text{s}/\pi r_0^2$. The source current gives rise to an electromagnetic field, which induces eddy currents in region $\Omega_\text{III}$. Since the source current is harmonic, we can solve the magnetoquasistatic PDE \eqref{eq:magquasistat} in the frequency domain.
The current $\vec{I}_\text{s}$ only has a $z$-component, therefore, just the $H_\varphi$ component of the magnetic field will be non-zero and with $\mu \vec{H}= \nabla \times\vec{A}$, the vector potential $\vec{A}$ can be reduced to $\vec{A}=A_z \vec{e}_z$. Moreover, the problem is translation-invariant and symmetric with respect to a rotation, thus we know that the magnetic field is given by $\vec{H}(r)=H_\varphi(r)\vec{e}_\varphi$.

The radius $r_1$, as well as the relative permeability $\mu_\text{r}$ and the current $I_\text{s}$, are modelled as uniform and independent random variables. The realizations of the random variables read
\begin{alignat}{2}
r_1(\randV)&=\bar{r}_1+Y_1(\randV), \,\,\,\,\,\,\,\,\,\,\,\,\,\,&&Y_1\sim\mathcal{U}(-2.54\,\text{mm}, 2.54\,\text{mm}), \label{eq:boundaryR1}\\
I_\text{s}(\randV)&=\bar{I}_\text{s}+Y_2(\randV), &&Y_2\sim\mathcal{U}(-10\,\text{A}, 10 \,\text{A}), \\
\mu_\text{r}(\randV)&=\bar{\mu}_\text{r}+Y_3(\randV), &&Y_3\sim\mathcal{U}(-400, 400),
\end{alignat}
with nominal values
\begin{alignat}{2}
\bar{r}_1 &= \mean[r_1] &&= 12.7 \, \text{mm}, \nonumber\\
\bar{I}_\text{s} &= \mean[I_\text{s}] &&= 100 \, \text{A}, \nonumber\\
\bar{\mu}_\text{r} &= \mean[\mu_\text{r}] &&= 1000.\nonumber
\end{alignat}
Table \ref{tab:tubeMaterial} summarizes the parameters describing the cable.

\begin{table}[t]
	\ra{1.3}
		\centering
		\caption{Parameters of the coaxial cable}
		\raisebox{\depth}{\begin{tabular}[htbp]{c|c||c|c}
		\toprule
		$\mu_\text{I}$  & $\mu_0$ & $r_0$ & $\SI{2.54}{\milli\meter}$\\
		$\mu_\text{II}$ & $\mu_0$ & $r_2$ & $\SI{25.4}{\milli\meter}$\\ \cline{3-4}
		$\mu_\text{III}$& $\mu_0\mu_{\text{r}}$ & $\bar{r}_1$ & $\SI{12.7}{\milli\meter}$ \\
		$\sigma_\text{I}$  & $\SI{0}{\mega\siemens\per\meter}$& $\bar{I}_\text{s}$ & $\SI{100}{\ampere}$\\
		$\sigma_\text{II}$ & $\SI{0}{\mega\siemens\per\meter}$& $\bar{\mu}_\text{r} $ & 1000 \\
		$\sigma_\text{III}$& $\SI{58}{\mega\siemens\per\meter}$& & \\
		\bottomrule
		\end{tabular}}
		\label{tab:tubeMaterial}
\end{table}

The solution of the problem is numerically obtained by solving the FE system~\eqref{eq:discrete_curlcurl}. It yields an approximation of the magnetic energy as defined in \eqref{eq:magneticEnergyGeneral}. Using a mesh sequence satisfying \eqref{eq:hLevel} the MLMC method is applied to estimate the mean value $\mean[W]$.
The mean value is also estimated using the stochastic collocation approach, which is well suited for this problem since only a low number of random variables are considered and the closed form solution is smooth \cite{Stoll_1974aa}. The evaluated mean energy is used as a reference and to determine the finest level $L$ in the MLMC algorithm.

\subsection{Permanent magnet synchronous machine}
The machine under consideration is based on the machine presented in~\cite{Pahner_1998aa}, which is a three-phase six pole PMSM (Fig.~\ref{fig:pmsm}). The stator has two slots per pole and per phase and a conventional distributed double-layer winding is used. The six rare-earth magnets in the rotor are buried. The rotor and stator are constructed from laminated steel. The stator consists of 36 teeth, which we assume to be welded together.

\begin{figure}
\centering
\def\svgwidth{1.0\columnwidth}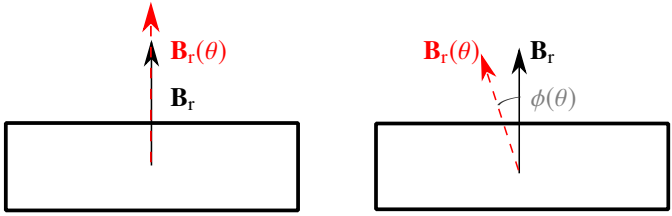
\caption{\label{fig:unc_magnets} Schematic view for the modelling of the magnet uncertainties, where on the left hand side the magnitude of the field is uncertain. On the right side the orientation of the field is uncertain.}
\end{figure}

The considered uncertainties in the PMSM are inherent in the teeth of the stator and in the permanent magnets in the rotor. Due to a welding process which causes material contamination, the reluctivities $\nu$ of the welding regions are considered random. This is also the case for the magnitudes of the remanent magnetic field ($B_\text{r}=|\vec{B}_\text{r}|$) of the magnets. Furthermore, due to the manufacturing process, the angle of the permanent magnets can be considered random as well as depicted in Fig.~\ref{fig:unc_magnets}. The modelling of those uncertainties is partially based on~\cite{Offermann_2013aa}. In total we have to deal with $48$ uncertain parameters.
The random parameters are assumed to be independently and uniformly distributed:
\begin{alignat}{2}
\nu_i(\randV)&=\bar{\nu}_i+Y_i(\randV), \,\,\,\,\,\,\,\,&&Y_i\sim\mathcal{U}(\SI{-330}{\meter\per\henry},\SI{330}{\meter\per\henry}), \\
B_{\text{r},j}(\randV)&=\bar{B}_{\text{r},j}+Y_j(\randV), &&Y_i\sim\mathcal{U}(\SI{-0.05}{\tesla},\SI{0.05}{\tesla}), \\
\phi_j(\randV)&=\bar{\phi}_j+Y_j(\randV), &&Y_j\sim\mathcal{U}(\SI{-3}{\degree},\SI{3}{\degree}),
\end{alignat}
with nominal values
\begin{alignat}{3}
\bar{\nu}_i & = \mean[\nu_i] && =  \SI{1100}{\meter\per\henry}\,\,\,&& \mathrm{where~} i \in \{1, \dots, 36\},\\
\bar{B}_{\text{r},j} & = \mean[B_{\text{r},j}] && = \SI{0.94}{\tesla} && \mathrm{where~} j \in \{1, \dots, 6\},\\
\bar{\phi}_j & =  \mean[\phi_{j}] && =\SI{0}{\degree} && \mathrm{where~} j \in \{1, \dots, 6\}.
\end{alignat}
As we disregard eddy-currents in this model, we can plug the uncertainties into \eqref{eq:magstatic} to obtain the stochastic problem. The QoI is again the total magnetic energy.

%% file: fig_tube_1.pdf_tex
\begingroup%
  \makeatletter%
  \providecommand\color[2][]{%
    \errmessage{(Inkscape) Color is used for the text in Inkscape, but the package 'color.sty' is not loaded}%
    \renewcommand\color[2][]{}%
  }%
  \providecommand\transparent[1]{%
    \errmessage{(Inkscape) Transparency is used (non-zero) for the text in Inkscape, but the package 'transparent.sty' is not loaded}%
    \renewcommand\transparent[1]{}%
  }%
  \providecommand\rotatebox[2]{#2}%
  \ifx\svgwidth\undefined%
    \setlength{\unitlength}{349.96875bp}%
    \ifx\svgscale\undefined%
      \relax%
    \else%
      \setlength{\unitlength}{\unitlength * \real{\svgscale}}%
    \fi%
  \else%
    \setlength{\unitlength}{\svgwidth}%
  \fi%
  \global\let\svgwidth\undefined%
  \global\let\svgscale\undefined%
  \makeatother%
  \begin{picture}(1,1)%
    \put(0,0){\includegraphics[width=\unitlength]{fig_tube_1.pdf}}%
    \put(0.71066676,0.80446837){\color[rgb]{0,0,0}\makebox(0,0)[lb]{\smash{$\Omega_{\mathrm{III}}$}}}%
    \put(0.64071297,0.62799408){\color[rgb]{0,0,0}\makebox(0,0)[lb]{\smash{$\Omega_{\mathrm{II}}$}}}%
    \put(0.53,0.50398514){\color[rgb]{0,0,0}\makebox(0,0)[lb]{\smash{$\Omega_{\mathrm{I}}$}}}%
    \put(0.42,0.67){\color[rgb]{0,0,0}\makebox(0,0)[lb]{\smash{$r_1$}}}%
    \put(0.48490686,0.4085936){\color[rgb]{0,0,0}\makebox(0,0)[lb]{\smash{$r_0$}}}%
    \put(0.23016653,0.58279382){\color[rgb]{0,0,0}\makebox(0,0)[lb]{\smash{$r_2$}}}%
  \end{picture}%
\endgroup%

%% file: uncertain_magnets.pdf_tex
\begingroup%
  \makeatletter%
  \providecommand\color[2][]{%
    \errmessage{(Inkscape) Color is used for the text in Inkscape, but the package 'color.sty' is not loaded}%
    \renewcommand\color[2][]{}%
  }%
  \providecommand\transparent[1]{%
    \errmessage{(Inkscape) Transparency is used (non-zero) for the text in Inkscape, but the package 'transparent.sty' is not loaded}%
    \renewcommand\transparent[1]{}%
  }%
  \providecommand\rotatebox[2]{#2}%
  \ifx\svgwidth\undefined%
    \setlength{\unitlength}{413.20371094bp}%
    \ifx\svgscale\undefined%
      \relax%
    \else%
      \setlength{\unitlength}{\unitlength * \real{\svgscale}}%
    \fi%
  \else%
    \setlength{\unitlength}{\svgwidth}%
  \fi%
  \global\let\svgwidth\undefined%
  \global\let\svgscale\undefined%
  \makeatother%
  \begin{picture}(1,0.26536126)%
    \put(0,0){\includegraphics[width=\unitlength]{uncertain_magnets.pdf}}%
    \put(0.79,0.23){\color[rgb]{0,0,0}\makebox(0,0)[lb]{\smash{$\vec{B}_\textrm{r}$}}}%
    \put(0.79,0.16){\color[rgb]{0,0,0}\makebox(0,0)[lb]{\smash{$\textcolor{gray}{\phi(\theta)}$}}}%
    \put(0.25,0.16){\color[rgb]{0,0,0}\makebox(0,0)[lb]{\smash{$\vec{B}_\textrm{r}$}}}%
    \put(0.25,0.23){\color[rgb]{0,0,0}\makebox(0,0)[lb]{\smash{$\textcolor{red}{\vec{B}_\textrm{r}(\theta)}$}}}%
    \put(0.63,0.23){\color[rgb]{0,0,0}\makebox(0,0)[lb]{\smash{$\textcolor{red}{\vec{B}_\textrm{r}(\theta)}$}}}%
  \end{picture}%
\endgroup%

%% file: Results.tex
\section{Results}
\label{sec:results}
In this section numerical results for the coaxial cable and the PMSM are presented and discussed.  
\subsection{Coaxial cable}
For this example, a closed-form solution is available, which is used to determine the finest level $L$, required to satisfy a pre-defined accuracy level. Based on this result, the use of the Richardson error indicator is verified. Finally, it is also investigated whether the use of nested or non-nested meshes has an impact on results of the MLMC method.

\subsubsection{Determining MLMC constants}
\begin{figure}[h!]   
  \centering
\begin{tikzpicture}
\begin{axis}[legend entries={$|\mean[W_\ell-W]|$,$\mathcal{O}(h^{2})$},
             grid=major,
             xlabel={$1/h$ in $\si{\per\meter}$},
             xmode=log,
             ymode=log,
             legend pos=south west,
             ylabel={$|\mean[W_\ell-W]|$ in $\si{\joule}$}]
\addplot[blue, mark=*, mark options={solid}, thick] table [ x expr={(\thisrowno{0}^-1)}, y index = {3}, col sep=comma]{constant_alpha_gpC.csv};
\addplot[black, thick, dashed] table [ x expr={(\thisrowno{0}^-1)}, y index = {4}, col sep=comma]{constant_alpha_gpC.csv};
\end{axis}
\end{tikzpicture}
\caption{FEM error in mean value for the coaxial cable (blue) and least square regression (black), indicating a weak error decay of $\alpha=2$.}
\label{fig:alpha_Coax}
\end{figure}
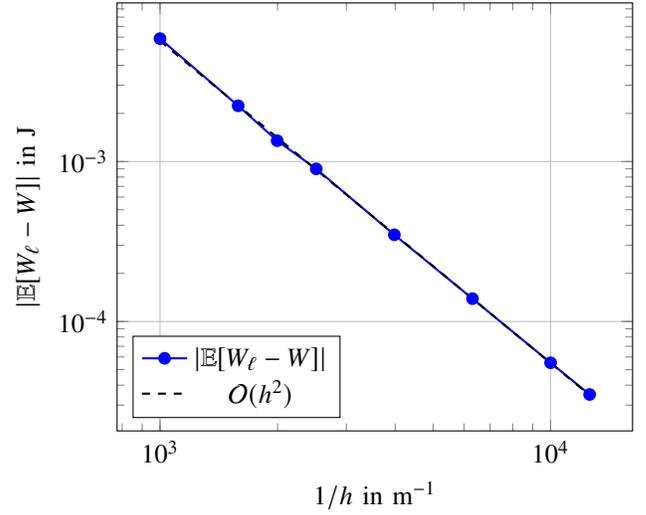

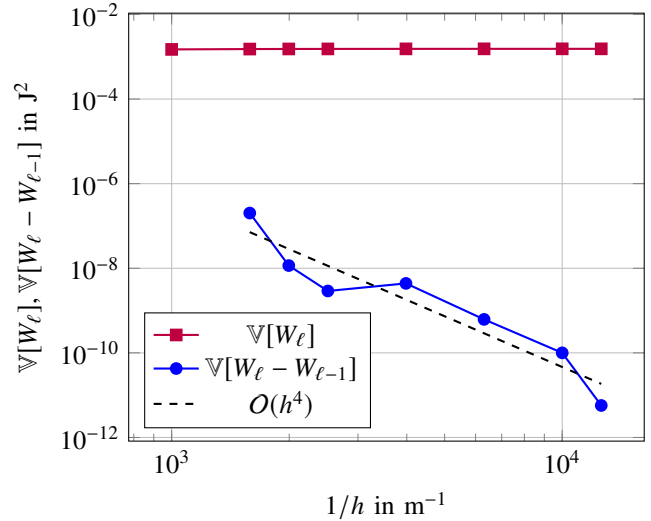
\begin{figure}[h!]     
  \centering
\begin{tikzpicture}
\begin{axis}[legend entries={$\var[W_\ell]$,$\var[W_\ell - W_{\ell-1}]$,$\mathcal{O}(h^{4})$},
             grid=major,
             xlabel={$1/h$ in $\si{\per\meter}$},
             xmode=log,
             ymode=log,
             legend pos=south west,
             ylabel={$\var[W_\ell]$,\,$\var[W_\ell - W_{\ell-1}]$ in $\si{\square\joule}$ }]
\addplot[purple, mark=square*, mark options={solid}, thick] table [ x expr={(\thisrowno{0}^-1)}, y index = {1}, col sep=comma]{constant_alpha_gpC.csv};
\addplot[blue, mark=*, mark options={solid}, thick] table [ x expr={(\thisrowno{0}^-1)}, y index = {1}, col sep=comma]{constant_beta_gpC.csv};
\addplot[black, thick, dashed] table [ x expr={(\thisrowno{0}^-1)}, y index = {2}, col sep=comma]{constant_beta_gpC.csv};
\end{axis}
\end{tikzpicture}
\caption{In red $\var[W_\ell]$ and in blue $\var[W_\ell - W_{\ell-1}]$ are plotted as function of $h_\ell$. A least square regression is shown in black, yielding in $\beta=4$ for the variance decay.}
\label{fig:beta_Coax}
\end{figure}

Before carrying out the MLMC analysis, we determine the parameters $\alpha,\beta$ of Theorem~\ref{th:mlmcTheorem} numerically. The quantities $\mean[W_\ell], \mean[W]$ and $\var[V_\ell]$ and $\var[V_\ell - V_{\ell-1}]$ are estimated using the stochastic collocation method. This is possible, since only three random inputs are present and the solution is a smooth function of these inputs.
The degree of the collocation method is chosen such that the relative error of the mean value is below $10^{-14}$, resulting in a polynomial degree of $p=12$. 
Applying a least square regression yields $\alpha=2$ and $\beta=4$, see Fig.~\ref{fig:alpha_Coax} and Fig.~\ref{fig:beta_Coax}. This is in agreement with the theoretical predictions in~\cite{ref:Teckentrup_FEM}.

It is noticeable that $\var[W_\ell]$ is nearly constant over all levels. This observation complies with the theoretical considerations, that the MSE can be divided into an error of the variance and an error caused by the FEM approximation. 

\subsubsection{Results based on the closed-form solution}
Figure \ref{fig:results_Coax} depicts $\meanML[W_L]$ over costs for different error bounds $\varepsilon$. The costs are determined according to \eqref{eq:costsGamma2}. The closed-form solution is plotted as a reference. Figure \ref{fig:samplesLvl_Coax} depicts the corresponding number of samples per level. Adding more levels reduces the weak error and one gets closer to the mean value determined by stochastic collocation. One clearly sees that the dominant costs originate in sampling on the coarsest level. This is expected due to the decreasing variance. Indeed, the number of samples calculated with \eqref{eq:NlBasic} is significantly smaller for levels $\ell=1, \ldots, L$, see Fig.~\ref{fig:beta_Coax}. Obviously, a higher accuracy requires more samples on every level and in some cases also an additional level, which leads to higher costs.
\begin{figure}[h!] 
\begin{tikzpicture}
\begin{axis}[legend entries={$\varepsilon=7 \cdot 10^{-5}$,$\varepsilon=5 \cdot 10^{-5}$, $\mean[W]$},
             grid=major,
             legend style={cells={align=left}},
             domain=7*10^7:2*10^9,
             xlabel={Costs},
             y tick label style={/pgf/number format/.cd,
            					fixed,
					            fixed zerofill,
					            precision=4,
						        /tikz/.cd
    							},		
             ylabel={$\meanML[W_L]$ in $\si{\joule}$},
             legend pos=south east,
             clip mode=individual]
\addplot[mark=*,mark options={solid},blue, very thick, 
         visualization depends on=\thisrow{alignment} \as \alignment,
         nodes near coords,
         point meta=explicit symbolic,
         every node near coord/.append style={anchor=\alignment},
         ] table [ x=costs, y=meanW, col sep=comma, meta index = 2]{results_7.0e-05.csv};
\addplot[mark=square*,mark options={solid},magenta, very thick] table [ x index = {0}, y index = {1}, col sep=comma]{results_5.0e-05.csv};
\addplot[color=red, very thick] { 0.196456911623244};
\end{axis}
\end{tikzpicture}
\caption{The mean value of the magnetic energy in the coaxial cable evaluated for different error bounds $\varepsilon$ with the closed-form solution to determine $L$.}
\label{fig:results_Coax}
\end{figure}
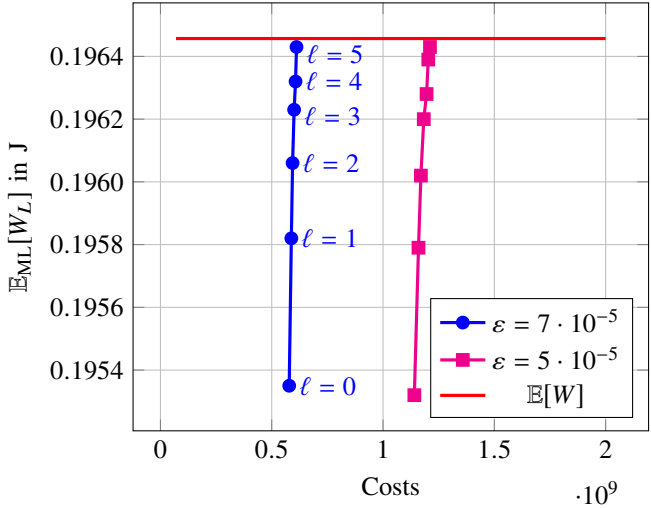
\begin{figure}[h]   
  \centering
\begin{tikzpicture}
\begin{axis}[legend entries={$\varepsilon=7 \cdot 10^{-5}$,$\varepsilon=5 \cdot 10^{-5}$},
             grid=major,
             legend style={cells={align=left}},
             xlabel={level},	
             ylabel={$N_\ell$},
             ymode=log,
             legend pos=north east,
             clip mode=individual]
\addplot[mark=*,mark options={solid},blue, very thick] table [ x index = {0}, y index = {1}, col sep=comma]{results_NumberOfSamples_7.0e-05.csv};
\addplot[mark=square*,mark options={solid},magenta, very thick] table [ x index = {0}, y index = {1}, col sep=comma]{results_NumberOfSamples_5.0e-05.csv};
\end{axis}
\end{tikzpicture}
\caption{Number of samples $N_\ell$ per level for modelling the coaxial cables. The circles indicate an error bound of $\varepsilon=5\cdot10^{-4}$, the squares a bound of $\varepsilon=7\cdot10^{-5}$. The closed-form solution was used to determine $L$.}
\label{fig:samplesLvl_Coax}
 \end{figure}
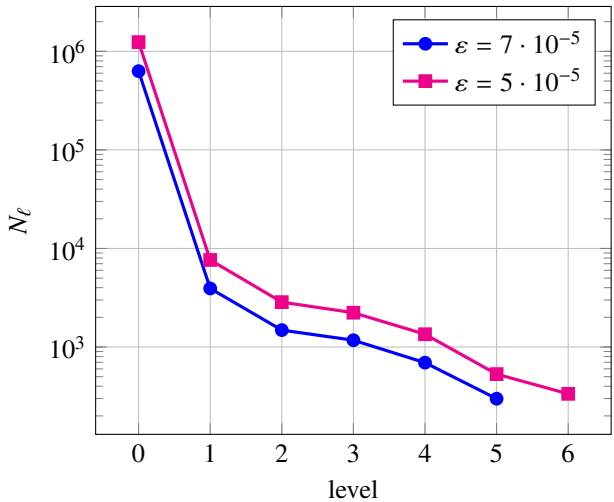

 \subsubsection{Richardson extrapolation}\label{subsec:richardsonExtrapol}
In general, we do not have a closed-form solution available. In this case, to determine the finest level $L$, we use Richardson extrapolation as outlined in Section \ref{sec:methodology}.
\begin{figure}[h]  
   \centering
 \begin{tikzpicture}
 \begin{axis}[legend entries={$\epsilon$, $\epsilon_\mathrm{Richardson}$,$\mathcal{O}(h^{2})$, $\mathcal{O}(h^{3})$},
              grid=major,
              xlabel={$1/h$ in $\si{\per\meter}$},
              xmode=log,
              ymode=log,
              ylabel={$\epsilon$,\,$\epsilon_\mathrm{Richardson}$ }]
 \addplot[mark=*,mark options={solid},green, very thick] table [ x index = {0}, y index = {1}, col sep=comma]{convWnom.csv};
 \addplot[mark=square*,mark options={solid},blue, very thick] table [ x index = {0}, y index = {1}, col sep=comma]{convWRichard.csv};
 \addplot[black, dashdotted, thick] table [ x index = {0}, y expr={(\thisrowno{0}^-2)*10^3.1}, col sep=comma]{convWnom.csv};
 \addplot[black, dashed, thick] table [ x index = {0}, y expr={(\thisrowno{0}^-3)*10^5.1}, col sep=comma]{convWnom.csv};
 \end{axis}
 \end{tikzpicture}
 \caption{Convergence of Richardson extrapolation and conventional FEM solution.}
 \label{fig:convRichard}
\end{figure}
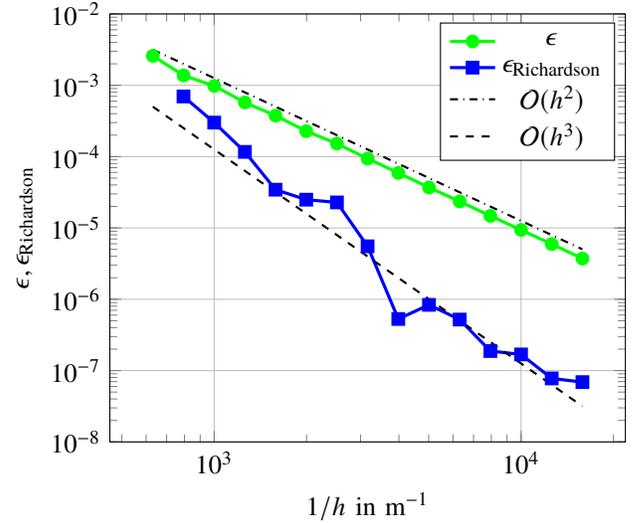
 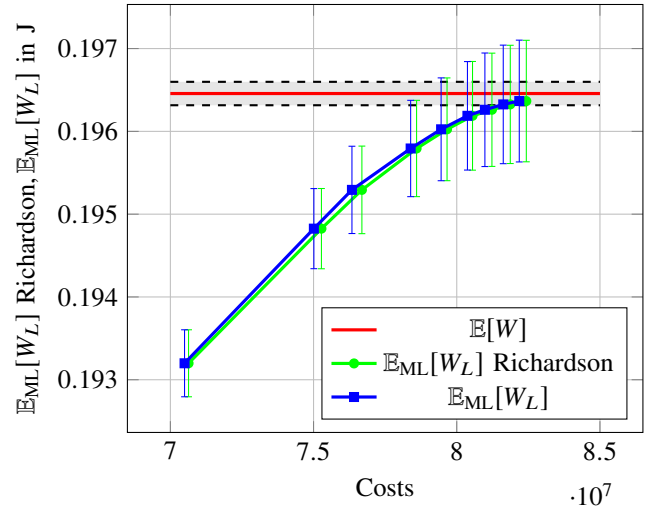
\begin{figure}[h]
    \centering
  \begin{tikzpicture}
  \begin{axis}[legend entries={$\mean[W]$,$\meanML[W_L]$ Richardson,$\meanML[W_L]$},
               grid=major,
               legend style={cells={align=left}},
               domain=0.7*10^8:0.85*10^8,
               xlabel={Costs},
               y tick label style={/pgf/number format/.cd,
              					fixed,
  					            fixed zerofill,
  					            precision=3,
  						        /tikz/.cd
      							},		
               ylabel={$\meanML[W_L]$ Richardson,\,$\meanML[W_L]$ in $\si{\joule}$},
               legend pos=south east,
               clip mode=individual]
                 \addplot[color=red, very thick] { 0.196456911623244};
                 \addplot[name path=upperLimit, color=black, thick, dashed, forget plot] { 0.196456911623244+sqrt(2)*10^-4};
                 \addplot[name path=lowerLimit, color=black, thick, dashed, forget plot] { 0.196456911623244-sqrt(2)*10^-4};
  \addplot[color=gray!20, forget plot]fill between[of=upperLimit and lowerLimit];
  \addplot[mark=*,mark options={solid},green,mark size=1.3pt, very thick,error bars/.cd, y dir=both, y explicit] table [         	x=costs_richardson,
	y=W_mean_richardson,
	y error plus expr=\thisrow{confidence_richardson},
    y error minus expr=\thisrow{confidence_richardson},
    col sep=comma,
    ]{richardsonVSanalyltic_eps_2e-4.csv};
  \addplot[mark=square*,mark options={solid},blue, mark size=1.3pt,very thick,error bars/.cd, y dir=both, y explicit] table [ x=costs_analytic, 
        y=W_mean_analytic, 
    	y=W_mean_richardson,
	y error plus expr=\thisrow{confidence_analytic},
    y error minus expr=\thisrow{confidence_analytic},
    col sep=comma,
    ]{richardsonVSanalyltic_eps_2e-4.csv};
  \end{axis}
  \end{tikzpicture}
  \caption{$\meanML[W_L]$ evaluated for error bound $\varepsilon=2 \cdot 10^{-4}$ with Richardson extrapolation and analytical solution. The error bars show the $\pm 3\sigma$ confidence interval. The red line depicts the reference solution obtained via the closed-form solution and the grey area indicates $\mean[W] \pm \varepsilon$.}
  \label{fig:richardsonVSanalytic}
        \end{figure}
 Figure \ref{fig:convRichard} depicts the convergence of the Richardson extrapolation and the conventional FEM solution. The error is defined as
 \begin{equation}
 \epsilon=\frac{\left|W_\ell(\boldsymbol{\mu_Y})-W(\boldsymbol{\mu_Y})\right|}{\left|W(\boldsymbol{\mu_Y})\right|},
 \end{equation}
 respectively 
 \begin{equation}
 \epsilon_\mathrm{Richardson}=\frac{\left|\richard{W}_\ell(\boldsymbol{\mu_Y})-W(\boldsymbol{\mu_Y})\right|}{\left|W(\boldsymbol{\mu_Y})\right|}.
 \end{equation}
 The results show the convergence gain, discussed in Section~\ref{sec:Rich_ex}.
 
 To verify the suitability of the Richardson extrapolation, the MLMC method was applied with Richardson extrapolation and with the analytical solution to estimate the weak error. The results for $\mean[W_L]$ over the cost $C$, based on both approaches, are shown in Fig.~\ref{fig:richardsonVSanalytic} for a user-specified accuracy ${\varepsilon=2 \cdot 10^{-4}}$. As a reference the closed-form solution $\mean[W]$ is plotted as well. Since we obtain probabilistic results, we consider confidence intervals in Fig.~\ref{fig:richardsonVSanalytic}. The central limit theorem \cite{ref:caflisch} implies that $\meanML[W_L]$ follows approximately a normal distribution, such that the shown $\pm 3 \sigma$ confidence intervals contain $99.7\%$ of the evaluations. The unknown standard deviation $\sigma$ is estimated with the MLMC results. This demonstrates that Richardson extrapolation can be used in absence of a closed-form solution in the MLMC framework. 

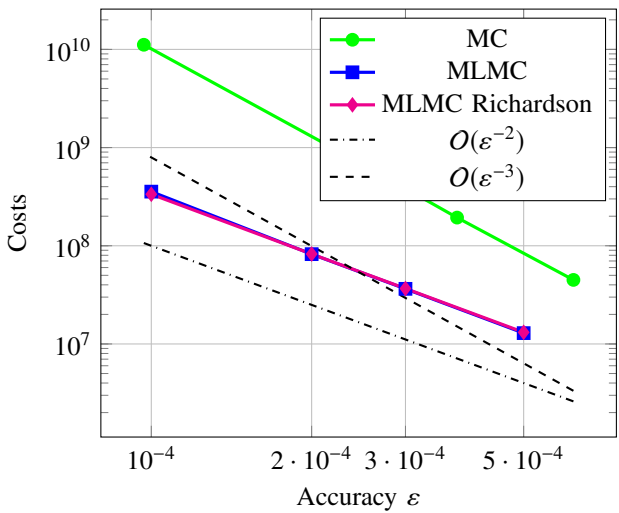
\begin{figure}[ht]   
    \centering
  \begin{tikzpicture}
  \begin{axis}[legend entries={MC, MLMC,MLMC Richardson,$\mathcal{O}(\varepsilon^{-2})$, $\mathcal{O}(\varepsilon^{-3})$},
               grid=major,
               xlabel={Accuracy $\varepsilon$},
               xtick={0.0001,0.0002,0.0003,0.0005},
               xticklabels={$10^{-4}$,$2\cdot10^{-4}$,$3\cdot10^{-4}$,$5\cdot 10^{-4}$},
               xmode=log,
               ymode=log,
               ylabel={Costs}]
  \addplot[mark=*,mark options={solid},green, very thick] table [ x index = {0}, y index = {1}, col sep=comma]{convMC.csv};
  \addplot[mark=square*,mark options={solid},blue, very thick] table [ x index = {0}, y index = {1}, col sep=comma]{convMLMCvsMLMCRichardson.csv};
  \addplot[mark=diamond*,mark options={solid},magenta, very thick] table [ x index = {2}, y index = {3}, col sep=comma]{convMLMCvsMLMCRichardson.csv};
  \addplot[black, dashdotted, thick] table [ x index = {0}, y expr={(\thisrowno{0}^-2)}, col sep=comma]{convMC.csv};
  \addplot[black, dashed, thick] table [ x index = {0}, y expr={(\thisrowno{0}^-3)*10^(-3.1)}, col sep=comma]{convMC.csv};
  \end{axis}
  \end{tikzpicture}
  \caption{Total costs of MLMC with the Richardson extrapolation, compared to MLMC where the exact solution is used to calculate $L$. The costs for the conventional MC method are plotted as well. All costs are related to the user-specified error bound $\varepsilon$.}
  \label{fig:totalCosts}
  \end{figure}

 Figure \ref{fig:totalCosts} compares the total costs of the MLMC method with and without Richardson extrapolation and the conventional MC method for a user-specified accuracy $\varepsilon$. The costs for the MLMC method are given by \eqref{eq:costsGamma2}.
 To determine the costs for conventional MC we have to analyse
 \begin{equation}
 \textrm{MSE}_{\textrm{MC}}= \underbrace{N^{-1}\var[W_\ell]}_{\textrm{MC error}} + \underbrace{\left(\mean[W_\ell]-\mean[W]\right)^2}_{\textrm{FE error}}.
 \end{equation}
 The upper bound for the MSE is $\varepsilon^2$, as it is for MLMC. The FE error is used to determine the required level and accordingly the mesh size $h_\ell$, so that the weak error is below $\varepsilon^2/2$. Once the required levels are determined, the number of samples can be identified by enforcing the MC error to be smaller than $\varepsilon^2/2$, leading to
 \begin{equation}
 N \ge \frac{2 \var[W_L]}{\varepsilon^2}.
 \end{equation}
The unknown variance in the previous relation can be approximated with its MC counterpart.  Then, the total costs can be estimated by $C^{\textrm{MC}}=N C^{\textrm{MC}}_\ell$, where $C^{\textrm{MC}}_\ell$ denotes the cost of one sample of $W_\ell$ on a mesh of size $h_\ell$.
  Since $\beta >  \gamma$, Theorem \ref{th:mlmcTheorem} predicts a computational complexity of $\highlight{C} \le c_4 \varepsilon^{-2}$, which is confirmed by the numerical results.
 It is clearly visible in Fig.~\ref{fig:totalCosts} that using the Richardson extrapolation produces similar results, with almost no additional costs.
   
\subsubsection{Nested mesh and remeshing}
To generate the levels for MLMC, we have to refine the mesh. This can either be accomplished by a nested refinement strategy or by generating a new mesh with smaller mesh size $h$. More details on level selection can be found in \cite{ref:Giles_2015aa}. Here, one level of nested refinement means that new mesh nodes are added at the center of the edges of the previous level without taking the original geometry into account. Since our geometry is a circle, the coarsest mesh defines the approximation quality of the radii. Another drawback is that the refinement factor $\Delta$ in $h_\ell=h_0 \Delta^\ell$ is is not freely adjustable. 
For a 2D mesh, at least an increase of the elements by factor four is obtained, which may quickly result in very dense meshes.
However, the domain $\Omega$ and the material distribution is technically different for each level and thus any sample of the reluctivity $\nu(\vec{x},\randV)$ is different for each level. Thus, a remeshing strategy violates the current theoretical MLMC framework. Nevertheless, our numerical results indicate that this effect has negligible influence and remeshing can and should be used in the MLMC method. 

\begin{figure}[t] 
  \begin{tikzpicture}
  \begin{axis}[legend entries={$\mean[W]$, $\meanML[W_L]${,}\\ remeshed,$\meanML[W_L]${,}\,nested},
               grid=major,
               legend style={cells={align=left}},
               domain=3.1*10^8:4.3*10^8,
               xlabel={Costs},
               y tick label style={/pgf/number format/.cd,
              					fixed,
  			            fixed zerofill,
  			            precision=3,
  				        /tikz/.cd
      							},		
               ylabel={$\mean[W]$,\,$\meanMC[W_L]$ in $\si{\joule}$},
               legend pos=south east,
               clip mode=individual]
                 \addplot[color=red, very thick] { 0.196456911623244};
                 \addplot[name path=upperLimit, color=black, thick, dashed, forget plot] { 0.196456911623244+10^-4};
                 \addplot[name path=lowerLimit, color=black, thick, dashed, forget plot] { 0.196456911623244-10^-4};
  \addplot[color=gray!20, forget plot]fill between[of=upperLimit and lowerLimit];
  \addplot+[mark=*,mark options={solid},green,mark size=1.3pt, very thick,error bars/.cd, y dir=both, y explicit] table [
                         x=costs,
                         y=meanW, 
                         y error plus expr=\thisrow{confidence},
                         y error minus expr=\thisrow{confidence},
                         col sep=comma,
                         ]{results_regular_1.0e-04_2.csv};
  \addplot[mark=square*,mark options={solid},mark size=1.3pt,blue, very thick, error bars/.cd, y dir=both, y explicit] table [
                         x=costs,
                         y=meanW, 
                         y error plus expr=\thisrow{confidence},
                         y error minus expr=\thisrow{confidence},
                         col sep=comma,
                         ]{results_nested_1.0e-04_2.csv};
  \end{axis}
  \end{tikzpicture}
  \caption{Comparing $\meanML[W_L]$ on remeshed and a nested meshes for $\varepsilon=10^{-4}$. The error bars show the $\pm 3\sigma$ confidence interval. The closed-form solution $W$ is used as a reference, the grey area indicates $\mean[W] \pm \varepsilon$.}
  \label{fig:NestedVSRemeshed}
  \end{figure}
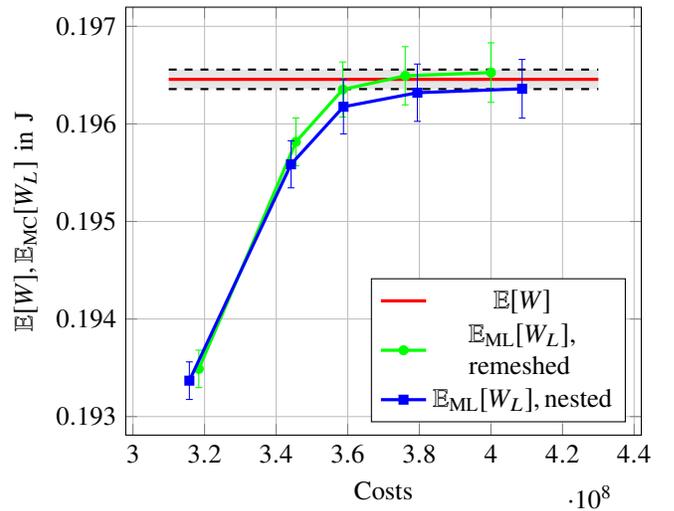

Figure \ref{fig:NestedVSRemeshed} shows the result for the magnetic energy calculated on nested and remeshed meshes for a user-specified accuracy $\varepsilon=10^{-4}$. The refined mesh is generated such that the DoFs per level are almost similar to the nested ones. The result of the closed-form solution is plotted as well.

In the following we want to estimate numerically the MSE given by
\begin{equation}
\textrm{MSE}=\mean\left[\left(\meanML[W_L]-\mean[W]\right)^2\right] \le \varepsilon^{2}.
\label{eq:MSE_MC}
\end{equation}
Therefore, the MLMC simulation is repeated $10$ times and the outer expectation value is estimated via MC. This is done for the nested mesh and the remeshing strategy described above. Table \ref{tab:MSE} shows the respective MSEs. Both approaches satisfy the conditions $\textrm{MSE} < \varepsilon^2$, thus we have demonstrated numerically that remeshing can be used for MLMC as well.
\begin{table}[t]
\centering
\caption{MSE for nested and remeshed levels. The expectation value $\meanML[W_L]$ was evaluated for $\varepsilon=10^{-4}$. The MSE was determined based on $10$ samples.} 
	\ra{1.3}
		\centering
		\raisebox{\depth}{\begin{tabular}[htbp]{@{}ll@{}}
		\toprule
 		mesh & MSE  \\
		\midrule
		nested & $\SI{1.3101 e-9}{\square\joule} $ \\
		remeshed & $\SI{5.1589 e-9}{\square\joule} $ \\ 
		\bottomrule
		\end{tabular}}
		\label{tab:MSE}
\end{table}

\subsection{PMSM}
We start our analysis of the MLMC method applied to the PMSM by estimating the constants $\alpha$ and $\beta$  numerically. After that we perform the MLMC simulation for different error bounds $\varepsilon$, discuss the results and compare the costs to conventional MC.

\subsubsection{Determining MLMC constants}
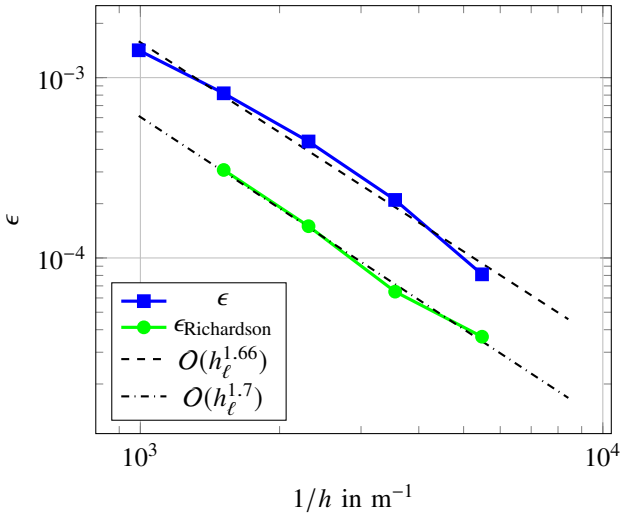
\begin{figure}[ht]
  \centering
  \begin{tikzpicture}
\begin{axis}[legend entries={$\epsilon$,$\epsilon_{\textrm{Richardson}}$,$\mathcal{O}(h_\ell^{1.66})$,$\mathcal{O}(h_\ell^{1.7})$},
             grid=major,
             xlabel={$1/h$ in $\si{\per\meter}$},
             xmode=log,
             ymode=log,
             legend pos= south west,
             ylabel={$\epsilon$}]
\addplot[mark=square*,mark options={solid},blue, very thick] table [ x=hMax fem, y=convergence fem, col sep=comma]{convergence_W_FEM.csv};
\addplot[mark=*,mark options={solid},green, very thick] table [ x=hMax richardson, y=convergence richardson, col sep=comma]{convergence_W_Richardson.csv};
\addplot[black, dashed, thick] table [ x=hMax, y=least square fem, col sep=comma]{convergence_W_least_square.csv};
\addplot[black, dashdotted, thick] table [ x=hMax, y=least square richard, col sep=comma]{convergence_W_least_square.csv};
\end{axis}
\end{tikzpicture}
\caption{Relative convergence of the energy $W$ over $1/h$. The linear least square fit $C h_\ell^{k}$, estimates a convergence rate of $\mathcal{O}(h_\ell^{1.66})$. The convergence for the Richardson extrapolation is given by $\mathcal{O}(h_\ell^{1.7})$.} 
\label{fig:convPMSMEnergyOverDoF}
  \end{figure}

To estimate $\alpha$ we study the convergence of the deterministic PDE. Therefore the relative deviation
\begin{equation}
\epsilon = \frac{|W_\ell(\boldsymbol{\mu_Y})-W_L(\boldsymbol{\mu_Y})|}{|W_L(\boldsymbol{\mu_Y})|} = \mathcal{O}\left(h_\ell^\alpha \right),
\end{equation}
with $\ell=0\ldots L-1$ is evaluated with nominal input values. In addition to the FE solution, the convergence order for the Richardson indicator $\epsilon_\textrm{Richardson}$ is computed. The results show a convergence rate of $\mathcal{O}\left(h_\ell^{1.66}\right)$ (Fig.~\ref{fig:convPMSMEnergyOverDoF}). The convergence for the Richardson extrapolation is given by $\mathcal{O}\left(h_\ell^{1.7}\right)$. A suboptimal convergence rate is to be expected due to the singular right-hand-side as discussed above.

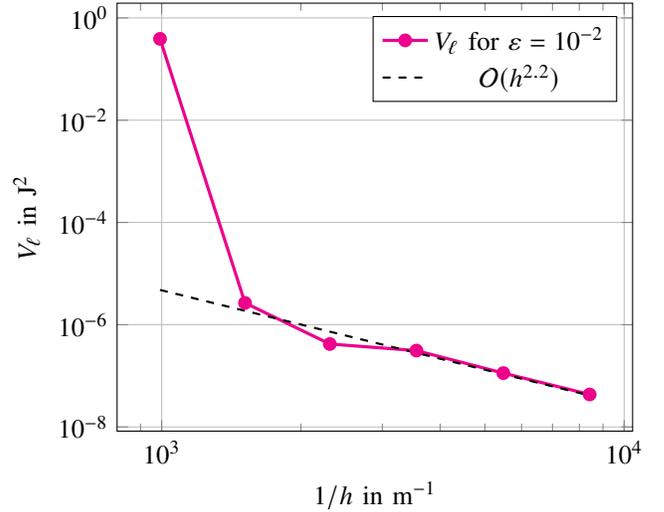
\begin{figure}[t]
  \centering
\begin{tikzpicture}
\begin{axis}[legend entries={$V_\ell$ for $\varepsilon=10^{-2}$, $\mathcal{O}(h^{2.2})$},
             grid=major,
             ymode=log,
             xmode=log,
             legend style={cells={align=left}},
             xlabel={$1/h$ in $\si{\per\meter}$},	
             ylabel={$V_\ell$ in $\si{\square\joule}$},
             legend pos=north east]
\addplot[mark=*,mark options={solid},magenta, very thick] table [ x expr={\thisrowno{9}^(-1)}, y index = {5}, col sep=comma]{results_ref_beta_5.0e-03.csv};
\addplot[dashed, black, thick] table [ x expr={\thisrowno{9}^(-1)}, y index = {7}, col sep=comma]{results_ref_beta_5.0e-03.csv};
\end{axis}
\end{tikzpicture}
\caption{Variance $V_0=\var[W_0]$, respectively $V_\ell=\var[W_\ell-W_{\ell-1}]$ evaluated with a set of samples $N_0=33641$, $N_1=N_2=1000$, $N_3=500$, $N_4=250$ and $N_5=125$ plotted over mesh size $h_\ell$. The dashed line shows the linear least square regression, where we get $\beta \approx 2.2$.}
\label{fig:convVariancePMSM}
\end{figure}

To estimate the constant $\beta$ we perform a MLMC simulation with additional samples, compared to the samples calculated with \eqref{eq:NlBasic}. The variances $V_0= \var[W_0]$, respectively ${V_\ell = \var[W_\ell-W_{\ell-1}]}$, are evaluated with a set of samples ${N_0=33641}$, ${N_1=N_2=1000}$, $N_3=500$, $N_4=250$ and $N_5=125$. The results are shown in Fig.~\ref{fig:convVariancePMSM} and a least square regression leads to $\beta \approx 2.2$. 

\subsubsection{MLMC results on PMSM}
The MLMC simulation is now performed for different error bounds $\varepsilon$. We obtain a mean magnetic energy of $\meanML[W_L]=\SI{24.697}{\joule}$ and a variance $\var[W_L]=\SI{0.390}{\square\joule}$. For nominal input data, the energy is given by $W=\SI{25.082}{\joule}$.

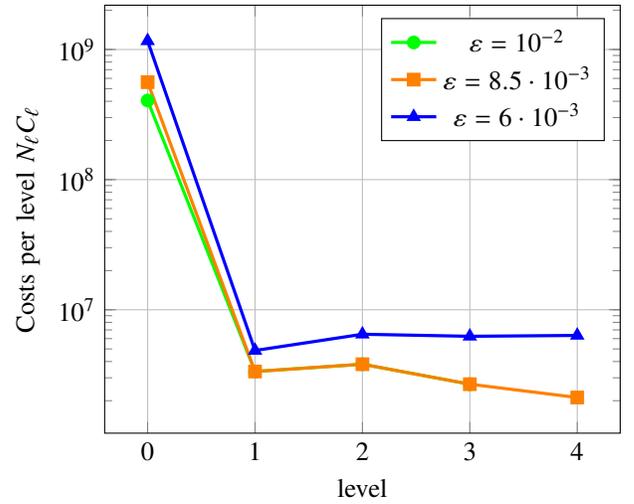
\begin{figure}[htb]   
  \centering
\begin{tikzpicture}
\begin{axis}[legend entries={$\varepsilon=10^{-2}$,$\varepsilon=8.5 \cdot 10^{-3}$,$\varepsilon=6 \cdot 10^{-3}$},
             grid=major,
             xtick={0,1,2,3,4,5},
             legend style={cells={align=left}},
             xlabel={level},	
             ymode=log,
             ylabel={Costs per level $N_\ell C_\ell$},
             legend pos=north east]
\addplot[mark=*,mark options={solid},green, very thick] table [ x index = {0}, y index = {6}, col sep=comma]{results_1_1.0e-02.csv};
\addplot[mark=square*,mark options={solid},orange, very thick] table [ x index = {0}, y index = {6}, col sep=comma]{results_1_8.5e-03.csv};
\addplot[mark=triangle*,mark options={solid},blue, very thick] table [ x index = {0}, y index = {6}, col sep=comma]{results_1_6.0e-03.csv};
\end{axis}
\end{tikzpicture}
\caption{The costs per level $N_\ell C_\ell$ plotted over the corresponding level for different error bounds $\varepsilon$. We can clearly see, that the dominant costs are an the coarsest level due to the variance reduction.}
\label{fig:CostsPerLevel}
\end{figure}

\begin{figure}[htb]   
  \centering
\begin{tikzpicture}
\begin{axis}[legend entries={MC,MLMC,$\mathcal{O}(\varepsilon^{-2})$,$\mathcal{O}(\varepsilon^{-3.5})$},
             grid=major,
             legend style={cells={align=left}},
             xlabel={Accuracy $\varepsilon$},	
             ymode=log,
             xmode=log,
             ylabel={Costs},
             legend pos=north east]
\addplot[mark=*,mark options={solid},green, very thick] table [ x index = {0}, y index = {1}, col sep=comma]{costs_MC.csv};
\addplot[mark=square*,mark options={solid},blue, very thick] table [ x index = {0}, y index = {1}, col sep=comma]{costs_MLMC.csv};
\addplot[black, dashdotted, thick] table [ x index = {0}, y index = {1}, col sep=comma]{costs_MLMC_least_square.csv};
\addplot[black, dotted, thick] table [ x index = {0}, y index = {1}, col sep=comma]{costs_MC_least_square.csv};
\end{axis}
\end{tikzpicture}
\caption{Total costs of MLMC with the Richardson extrapolation performed on the PMSM. The estimated costs for the conventional MC method are plotted as well. All costs are related to the user-specified error bound $\varepsilon$.}
\label{fig:totalCosts_PMSM}
\end{figure}
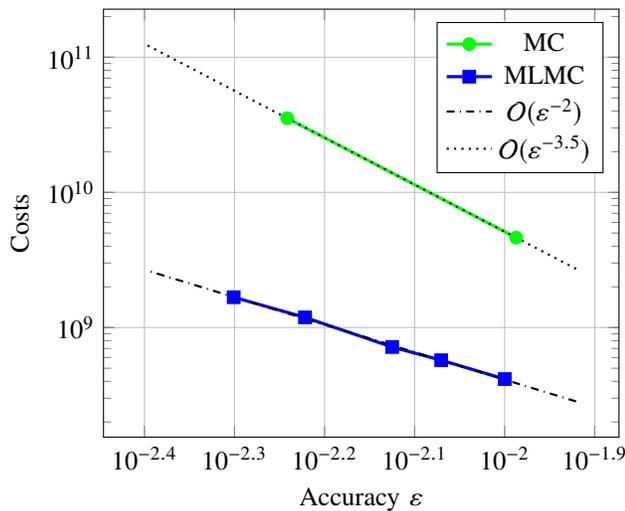

As expected, the variance of level $0$ is dominating, so are the costs (Fig.~\ref{fig:CostsPerLevel}).
The total costs for a MLMC and a conventional MC simulation for a user-specified accuracy $\varepsilon$ are compared in Fig.~\ref{fig:totalCosts_PMSM}. Since simulations with the conventional MC method become intractable for high accuracy, due to the high costs for one sample, the total costs are estimated. This is done in a similar way as in section~\ref{subsec:richardsonExtrapol}. The analytic solution $W$ in the FE error \eqref{eq:MSE_MC} is now replaced by the Richardson extrapolation. MLMC again outperforms conventional MC. The additional costs for the conventional MC are related to the Richardson extrapolation. 


%% file: Conclusion.tex
\section{Conclusion}
\label{sec:conclusion}
The Multilevel Monte Carlo method was successfully applied to an academic example and a real world problem. We have shown that the Richardson extrapolator is an appropriate indicator for the weak error and can be used to determine the finest level. Additionally, we have demonstrated that non-nested meshes can be used, which is essential as the number of degrees of freedom can quickly become very large when using nested meshes.
The numerical results show that MLMC drastically outperforms conventional MC. 




%